\documentclass[final,5p,times,twocolumn]{elsarticle}

\usepackage{amsmath,amssymb}
\usepackage{graphicx}
\usepackage{booktabs}
\usepackage{siunitx}
\usepackage{hyperref}
\hypersetup{
    pdftitle={Scalable operating currents in amorphous W-N/FePt superconducting rectifiers},
    pdfauthor={S. Russo, A. Sanson, Yeonkyu Lee, Jeehoon Kim, N. Haberkorn}
}

\biboptions{sort&compress}

\begin{document}
\begin{frontmatter}

\title{Scalable operating currents in amorphous W--N/FePt superconducting rectifiers}

\author[inst1]{S. Russo\fnref{equal}}
\author[inst1]{A. Sanson\fnref{equal}}
\author[inst3]{Yeonkyu Lee}
\author[inst3]{Jeehoon Kim}
\author[inst1,inst2]{N. Haberkorn\corref{cor1}}

\cortext[cor1]{Corresponding author.}
\ead{nestor.haberkorn@ib.edu.ar}

\fntext[equal]{S. Russo and A. Sanson contributed equally to this work.}

\affiliation[inst1]{
    organization={Instituto Balseiro, Universidad Nacional de Cuyo and Comisi\'on Nacional de Energ\'ia At\'omica},
    addressline={Av. Bustillo 9500},
    city={San Carlos de Bariloche},
    postcode={8400},
    country={Argentina}
}

\affiliation[inst2]{
    organization={Instituto de Nanociencia y Nanotecnolog\'ia (CNEA-CONICET), Nodo Bariloche},
    addressline={Av. Bustillo 9500},
    city={San Carlos de Bariloche},
    postcode={8400},
    state={R\'io Negro},
    country={Argentina}
}

\affiliation[inst3]{
    organization={Department of Physics, Pohang University of Science and Technology (POSTECH)},
    addressline={77 Cheongam-ro, Nam-gu},
    city={Pohang},
    postcode={37673},
    state={Gyeongsangbuk-do},
    country={Republic of Korea}
}

\begin{abstract}
We demonstrate high-efficiency, field-free superconducting diode behavior in W–N/FePt bilayers fabricated entirely by sputtering at room temperature. The devices combine amorphous W–N superconducting thin films with a 48-nm-thick FePt layer exhibiting remanent stripe-domain magnetization. We used lithographically patterned microbridges to compare W–N/FePt bilayers with single-layer W–N reference films at three nominal W–N thicknesses (18, 24, and 28 nm). The bilayers exhibit superconducting transition temperatures of approximately 4.1–4.3 K. At zero applied magnetic field, the current–voltage (I–V) characteristics exhibit pronounced nonreciprocity, controlled by the orientation of the in-plane remanent FePt magnetization relative to the transport current. The diode response is maximized for perpendicular alignment, reverses polarity upon magnetization reversal, and is strongly suppressed when the magnetization is aligned with the current. This behavior is consistent with asymmetric vortex entry and motion governed by the magnetic stray fields generated near the bridge edges. At 3 K with the remanent magnetization perpendicular to the current, increasing the W–N thickness from 18 to 28 nm widens the current interval between the dissipation thresholds for opposite polarities from about 0.3 to 1.55 mA. All three bilayers exhibit diode efficiencies of approximately 75–80$\%$ in this configuration.
\end{abstract}


\end{frontmatter}

\section{Introduction}
Diodes are fundamental building blocks of electronic circuits because their nonreciprocal current–voltage (I–V) characteristics enable current rectification and signal processing \cite{JiangHu2022, Nadeem2023}. Extending this functionality to superconducting circuits is attractive for low-dissipation cryogenic electronics. In a superconducting diode, the threshold for the onset of dissipation depends on the current polarity, resulting in unequal critical-current magnitudes, $|I_c^+|\neq |I_c^-|$. Achieving this response without an externally applied magnetic field during operation is particularly desirable for circuit integration. In conventional equilibrium superconducting systems, such nonreciprocity generally requires the appropriate breaking of both inversion and time-reversal symmetries. The superconducting diode effect has been demonstrated in Josephson junctions \cite{Golod2022}, kagome superconductors \cite{Le2024}, and various superconducting heterostructures \cite{Lin2022,Hou2023}. It can also arise from controlled vortex motion in geometrically engineered superconducting devices \cite{Lyu2021,Porrati2025} and superconductor/ferromagnet (SC/FM) hybrids \cite{Hou2023, Carapella2010, JiangHu2022, Zapata2026}.

In type-II superconducting devices, vortex entry and motion provide a natural route to polarity-dependent dissipation \cite{Vodolazov2005}. The current-induced Lorentz force drives vortex motion, while geometrical barriers, material defects, and engineered pinning landscapes determine the thresholds for flux entry and depinning \cite{Porrati2025,Carapella2010,Aladyshkin2010,Cerbu2013,Zapata2026}. Asymmetric device geometries and vortex-ratchet potentials can therefore generate nonreciprocal transport, often under an applied magnetic field. At zero applied field, a remanent magnetic configuration can provide an internal magnetic landscape that breaks time-reversal symmetry and, in combination with an appropriate spatial asymmetry, enables nonreciprocal vortex transport. In this context, SC/FM hybrids offer a relatively simple platform for realizing this functionality. Ferromagnetic layers can generate magnetic stray fields (MSFs) that modify vortex-entry barriers and guide vortex motion. Under suitable magnetic configurations and device geometries, these effects can produce strongly asymmetric dissipation thresholds and high-efficiency superconducting rectification without an externally applied field during operation \cite{Hou2023, Carapella2010,Zapata2026,Cruz2026,Zapata2026NbNFePt}. 

Our previous results indicate that the efficiency of SC/FM superconducting diodes is governed by the interplay between the magnetic landscape of the FM and the vortex dynamics of the superconducting layer \cite{Zapata2026,Cruz2026,Zapata2026NbNFePt}. Permalloy films close to the onset of stripe formation can provide a remanent dipolar-like magnetic configuration in the incipient-stripe regime, with more developed stripe structures emerging at larger thicknesses \cite{Carapella2010,Zapata2026,Voltan2016,Caputo2017}, whereas FePt develops well-defined stripe domains above a thickness of approximately 40 nm \cite{Alvarez2015,Blatter2024}. In FePt-based devices, the diode response strongly depends on the superconducting material. In NbN/FePt, dissipative vortex motion occurs in both current polarities before switching, while the substantial threshold of the lower-current branch suggests a stronger contribution from vortex pinning \cite{Zapata2026NbNFePt}. In contrast, disordered Mo$_2$N with weaker pinning displays a low-current dissipative branch together with an almost dissipationless high-current branch that switches abruptly at a much larger current, leading to substantially higher diode efficiencies \cite{Cruz2026}. These observations suggest that high-efficiency field-free vortex rectification can be favored by two complementary conditions: a magnetic texture that promotes asymmetric flux entry at the bridge edges while minimizing premature dissipation associated with vortices in the interior, and a superconducting layer with sufficiently weak pinning to allow flux injected in the dissipative polarity to move at low current. For the opposite polarity, suppression of inward flux injection may preserve a large low-dissipation current-carrying region and allow the critical current to approach a significant fraction of the depairing limit. This picture motivates the exploration of amorphous superconductors with potentially weak vortex pinning as a route toward high-efficiency superconducting rectification.

Here, we investigate nonreciprocal transport in bilayers comprising amorphous nitrogen-doped tungsten (W--N) and FePt, extending our previous studies on NbN/FePt and Mo$_2$N/FePt heterostructures. W--N was selected as an amorphous superconductor with a transition temperature ($T_c$) near 5 K and potentially weak vortex pinning, with films grown by reactive sputtering under conditions previously shown to suppress crystalline ordering \cite{Hofer2023}. A 48-nm-thick FePt layer was chosen to provide a stable remanent magnetic configuration characterized by nanoscale stripe domains, whose orientation can be controlled through the preceding in-plane saturation field \cite{Alvarez2015}. By comparing single-layer W--N microbridges and corresponding W--N/FePt bilayers with superconducting thicknesses of 18, 24, and 28 nm, we investigate whether the operating-current scale can be adjusted while preserving high diode efficiency. Field-dependent measurements on the single-layer bridges are used to assess possible contributions from geometric edge asymmetries \cite{Porrati2025}, while measurements under controlled remanent magnetic states establish the magnetic contribution to the nonreciprocal response. Finally, we examine the temperature dependence of the operating currents and distinguish between rectification associated with the onset of dissipation and that involving abrupt switching to a high-resistance state.

\section{Experimental methods}

W--N single layers and W--N/FePt bilayers were deposited on silicon substrates by sputtering at room temperature. Metallic W and FePt targets, each 0.038~m in diameter, were used. The chamber base pressure was below $3\times10^{-5}$~Pa. The W--N layers were grown by DC reactive sputtering at a power of 50~W and a target-to-substrate distance of 0.07~m. The total gas pressure (Ar + N$_2$) was maintained at 0.66~Pa, with N$_2$ accounting for 15\% of the gas mixture, following the conditions reported in Ref.~\cite{Hofer2023}. After W--N deposition, the chamber was evacuated to its base pressure before depositing FePt in pure Ar at 0.4~Pa and 20~W, with the same target-to-substrate distance. Three W--N single layers were prepared using deposition times of 60, 75, and 90~s, corresponding to nominal thicknesses of approximately 18, 24, and 28~nm, respectively. Three additional W--N layers with the same nominal thicknesses were capped with FePt deposited for 180~s, yielding a nominal FePt thickness of approximately 48~nm. Hereafter, the samples are denoted as W--N($d$) and W--N($d$)/FePt for the single layers and bilayers,
respectively, where $d$ is the nominal W--N thickness in nanometers.

Surface morphology and magnetic-domain structure were investigated by atomic force microscopy (AFM) and magnetic force microscopy (MFM), respectively, using a Park Systems NX10 microscope. Scanning electron microscopy (SEM) images of the microbridges, including details of their edges, were acquired using a Zeiss Gemini scanning electron microscope. Cross-sectional images of the W--N(28)/FePt bilayer were obtained using the same instrument operated in scanning transmission electron microscopy (STEM) mode.

Magnetization hysteresis loops were measured at 10 K using a commercial magnetometer (Cryogenic), with the magnetic field applied parallel to the film surface. Electrical transport measurements were performed on microbridges approximately 40~$\mu$m long ($L$) and 4.5~$\mu$m wide ($w$), using a standard four-terminal configuration. Magnetic fields of up to 1~T could be applied in two geometries. For measurements of the upper critical field and current--voltage (I--V) characteristics of the single-layer W--N films, the magnetic field was applied perpendicular to the film surface. For preparation of the remanent magnetic states used to characterize the diode effect in W--N/FePt bilayers, the field was applied within the film plane as reported elsewhere \cite{Zapata2026}. Its direction relative to the fixed microbridge was varied by rotating the electromagnet while keeping the sample stationary in the cryocooler. To prepare the remanent magnetic state, an in-plane field of 0.5~T, sufficient to saturate the FePt layer, was applied along a selected direction and subsequently removed. The I--V characteristics were then measured at zero applied magnetic field. The preparation angle $\theta$ specifies the direction of the in-plane preparation field relative to the positive current direction (Fig.~\ref{fig:5}). In this convention, $\theta=0^\circ$ and $180^\circ$ correspond to parallel and antiparallel orientations, respectively, while $\theta=90^\circ$ and $270^\circ$ correspond to the two opposite perpendicular orientations. Current--voltage curves were acquired with a Keithley 2182A nanovoltmeter and a Keithley 6221 current source operating in synchronized mode, using current pulses of 0.2~ms duration. The critical currents were determined using a voltage criterion of $|V|=2~\mu$V. For I--V characteristics exhibiting a gradual onset of dissipation, $I_c$ was obtained by extrapolating the approximately linear portion of the curves in a log--log representation to the selected voltage criterion. For abrupt transitions crossing this threshold, $I_c$ was identified with the switching current. The same procedure was applied to both current polarities. In addition, $I_c^*$ denotes the magnitude of the current at which the already dissipative branch switches abruptly to the normal-resistance state. The reproducibility of the transport response was verified using at least two microbridges, which exhibited quantitatively consistent critical currents and diode efficiencies under equivalent measurement conditions.

\section{Results and discussion}

\subsection{Structural, magnetic and superconducting properties}

The samples investigated in this work were designed following our previous results on nitrogen-doped W thin films, which showed sharp superconducting transitions with $T_c$ close to 5 K for W--N thicknesses above approximately 20 nm, together with extremely smooth surfaces favorable for the subsequent growth of homogeneous FePt layers \cite{Hofer2019, Hofer2023}. As discussed above, 48-nm-thick FePt films deposited at room temperature develop stripe-domain structures, providing the remanent magnetic configuration required for the field-free diode response. In the following, we first describe the microbridge geometry and the structural characteristics of the W--N/FePt heterostructures, and subsequently discuss the electrical properties of the single layers and bilayers together with the magnetic properties determined from magnetization and MFM measurements.

Figure~1(a) shows a representative microbridge with length $L = 40~\mu$m and width $w \approx 4.5~\mu$m, patterned by optical lithography. The bridge exhibits a highly uniform geometry along the active region. The inset shows a higher-magnification SEM image of a representative bridge edge, revealing a small nanoscale saw-tooth-like modulation with characteristic dimensions typically below 100 nm, comparable to edge imperfections previously observed in lithographically defined NbN microbridges \cite{Zapata2025MatLett}. To independently verify the layer thicknesses and the quality of the heterostructure, cross-sectional STEM measurements were performed on the W--N(28)/FePt sample. Figure~1(b) shows a representative cross section, revealing W--N and FePt thicknesses of approximately 29 and 48 nm, respectively, in good agreement with the deposition-rate calibrations. Both layers are laterally homogeneous and exhibit comparatively flat interfaces over the investigated region. The amorphous carbon layer visible above the FePt was deposited as a protective layer during cross-sectional sample preparation.

\begin{figure}[tbp]
\centering
 \includegraphics[width=8cm]{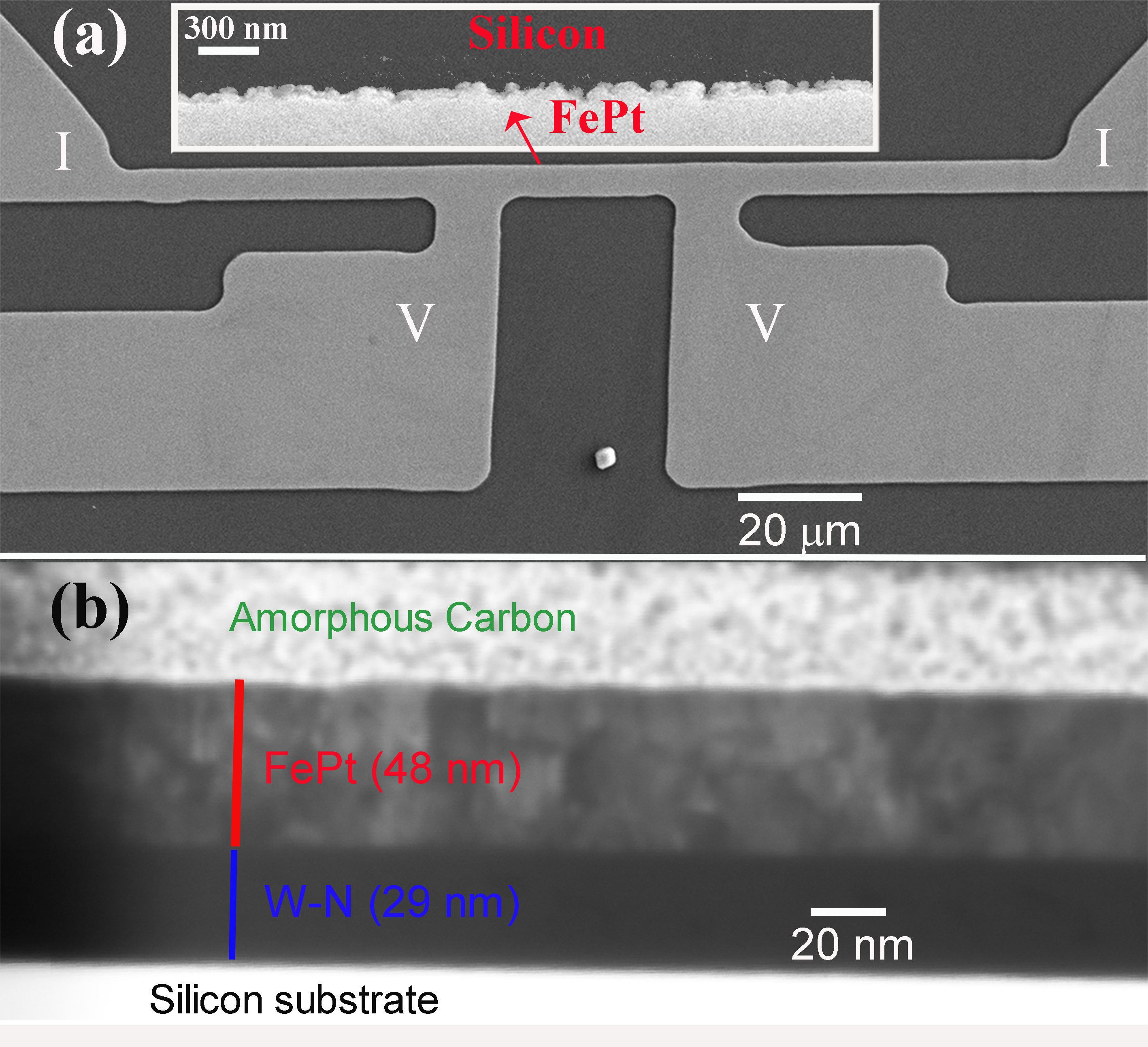}
\caption{Electron microscopy characterization of a representative W--N/FePt device. (a) SEM image of a patterned microbridge showing the current (I) and voltage (V) leads and the active bridge region. The inset shows a higher-magnification image of a representative bridge edge, revealing the nanoscale edge modulation produced during patterning. (b) Cross-sectional STEM image of the W--N(28)/FePt heterostructure, showing a W--N thickness of approximately 29 nm and a 48 nm FePt layer on the Si substrate. The amorphous carbon protective layer deposited during cross-sectional sample preparation is also indicated.}
\label{fig:1}
\end{figure}

Figure~2 shows the magnetic characterization of a representative W--N(18)/FePt bilayer. The in-plane magnetization loop measured at 10 K exhibits the characteristic response of a stripe-domain FePt film, with a saturation magnetization of approximately $1100~\mathrm{emu}\cdot\mathrm{cm}^{-3}$ and a coercive field $\mu_0H_c \approx 15$ mT. Upon reducing the field from saturation, the magnetization progressively departs from the fully in-plane configuration as the stripe-domain state develops, introducing a modulated out-of-plane component and consequently reducing the in-plane projection of the magnetization \cite{Alvarez2015,Voltan2016}. This produces the characteristic gradual evolution of $M(H)$ between saturation and remanence observed in the loop. Similar hysteresis curves were obtained for the other W--N/FePt bilayers, indicating that their magnetic response is mainly determined by the common 48-nm-thick FePt layer. The inset shows a room-temperature MFM image acquired in the remanent state after saturating the sample along the horizontal direction and subsequently removing the field. The image reveals a well-developed nanoscale stripe-domain pattern, preferentially aligned along the previous saturation direction, consistent with the magnetic configuration inferred from the hysteresis loop \cite{Alvarez2015}.

\begin{figure}[tbp]
\centering
\includegraphics[width=\linewidth]{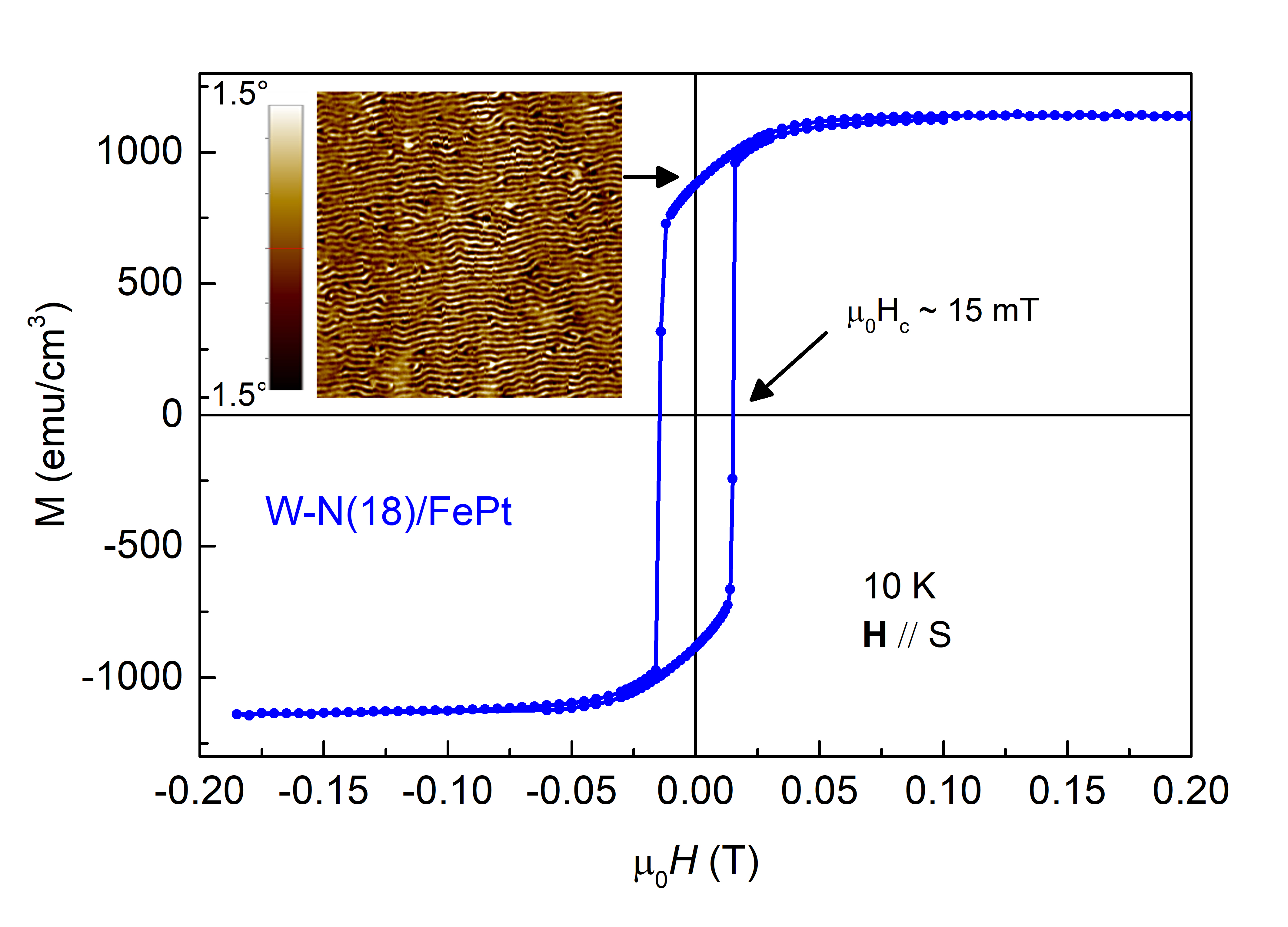}
\caption{Magnetic characterization of a representative W--N(18)/FePt bilayer. The main panel shows the $M(H)$ hysteresis loop measured at 10 K, with a saturation magnetization $M_s \approx$ $1100~\mathrm{emu}\cdot\mathrm{cm}^{-3}$ and a coercive field $\mu_0H_c \approx 15$ mT. The gradual reduction of the magnetization upon decreasing the field from saturation is characteristic of the formation of the stripe-domain state. The inset shows a room-temperature 5 x 5 $\mu$m$^2$ MFM phase image acquired in remanence after saturating the FePt layer along the horizontal direction and removing the applied field, revealing the nanoscale stripe-domain pattern characteristic of the 48-nm-thick FePt film. Similar hysteresis loops were obtained for the other W--N/FePt bilayers.}
\label{fig:2}
\end{figure}

The superconducting parameters of the single-layer W--N films were obtained from temperature- and magnetic-field-dependent transport measurements. Figure~3(a) shows the resistance normalized to its normal-state value, $R/R^N$, as a function of temperature for W--N(18), W--N(24), and W--N(28). All three films exhibit relatively sharp superconducting transitions, with $T_c$ values in the narrow range of approximately 4.7--4.8 K, defined using the criterion $R/R^N=0.8$. A weak increase in $T_c$ is observed with increasing W--N thickness, amounting to only about 0.1 K across the entire thickness range. The normal-state resistivities were estimated to be $130(10)$, $125(10)$, and $115(10)~\mu\Omega\cdot\mathrm{cm}$ for W--N(18), W--N(24), and W--N(28), respectively. Although a slight decrease in resistivity is observed with increasing film thickness, the values overlap within the experimental uncertainty and remain within the range expected for highly disordered W--N films. The inset of Fig.~3(a) shows the evolution of the superconducting transition of W--N(28) under magnetic fields up to approximately 1 T applied perpendicular to the film surface. Increasing field shifts the transition toward lower temperature and produces a low-resistance tail at the highest fields. This feature is associated with the small critical current of the micrometer-scale bridge close to the superconducting transition and does not significantly affect the determination of $T_c(H)$ using the 80\% resistance criterion. An approximately linear dependence of $\mu_0 H_{c2}$ on temperature was obtained close to $T_c$, with $\left|\mu_0 dH_{c2}/dT\right|_{T_c}$ ranging from 2.20 to 2.35 T K$^{-1}$ across the three films. Within the orbital dirty-limit Werthamer--Helfand--Hohenberg (WHH) approximation, $\mu_0 H_{c2}(0)$ was estimated as $\mu_0 H_{c2}(0)=-0.693\,T_c\left.\mu_0\frac{dH_{c2}}{dT}\right|_{T_c}$ \cite{Werthamer1966}. This extrapolation yields $\mu_0 H_{c2}(0)=7.6(1)$, $7.5(2)$, and $7.3(2)$ T for W--N(18), W--N(24), and W--N(28), respectively. The corresponding zero-temperature coherence lengths, calculated from $\xi(0)=[\Phi_0/(2\pi\mu_0 H_{c2}(0))]^{1/2}$, are approximately 6.6--6.7 nm. The normal-state electron diffusion coefficient was estimated from the slope near $T_c$ using $D=4k_B/\{\pi e|\mu_0\,dH_{c2}/dT|_{T_c}\}$, resulting in $D\approx0.5$ cm$^2$\,s$^{-1}$ \cite{Tinkham1996}. Figure~3(b) shows the corresponding normalized superconducting transitions after deposition of the 48-nm-thick FePt layer. In contrast to the nearly thickness-independent $T_c \approx 4.8$ K of the single-layer W--N films, the bilayers exhibit a clear suppression of superconductivity. For W--N(24)/FePt and W--N(28)/FePt, $T_c$ is reduced to approximately 4.2--4.3 K while the transitions remain relatively sharp. The effect becomes substantially stronger for W--N(18)/FePt, for which the transition shifts to lower temperature and broadens considerably. This enhanced suppression in the thinnest film is consistent with the increasing influence of the SC/FM proximity effect as the W--N thickness is reduced and approaches the regime in which superconductivity becomes strongly sensitive to the interface \cite{Buzdin2005}.

To establish an intrinsic current scale for comparison with the W--N/FePt bilayers, the depairing current of the single-layer films was estimated using the dirty-limit expression $I_{\mathrm{dep}}(0)=0.74w\Delta_0^{3/2}/[eR_s\sqrt{\hbar D}]$, where $R_s=\rho^N/d$ is the normal-state sheet resistance and $\Delta_0=1.76k_BT_c$ is the weak-coupling BCS gap \cite{ClemKogan2012}. Its temperature dependence was approximated as $I_{\mathrm{dep}}(T)=I_{\mathrm{dep}}(0)[1-(T/T_c)^2]^{3/2}$. For comparison among films of different thickness, the corresponding depairing current density was calculated as $J_{\mathrm{dep}}(T)=I_{\mathrm{dep}}(T)/(wd)$, which gives $J_{\mathrm{dep}}(0)=0.74\Delta_0^{3/2}/[e\rho^N\sqrt{\hbar D}]$. The resulting $J_{\mathrm{dep}}(0)$ values are approximately 6--7 MA cm$^{-2}$ for all three films and decrease to approximately 3 MA cm$^{-2}$ at 3 K. Because the intrinsic current-density scale remains similar across the series, the increase in W--N thickness translates directly into a substantial increase in the absolute depairing-current scale: $I_{\mathrm{dep}}(3\,\mathrm{K})$ increases from approximately 2.3 mA for W--N(18) to 3.2 mA for W--N(24) and 4.2 mA for W--N(28). The superconducting and normal-state parameters estimated for the three W--N films are summarized in Table~\ref{tab:WN_properties}.

\begin{figure}[t]
\centering
 \includegraphics[width=\linewidth]{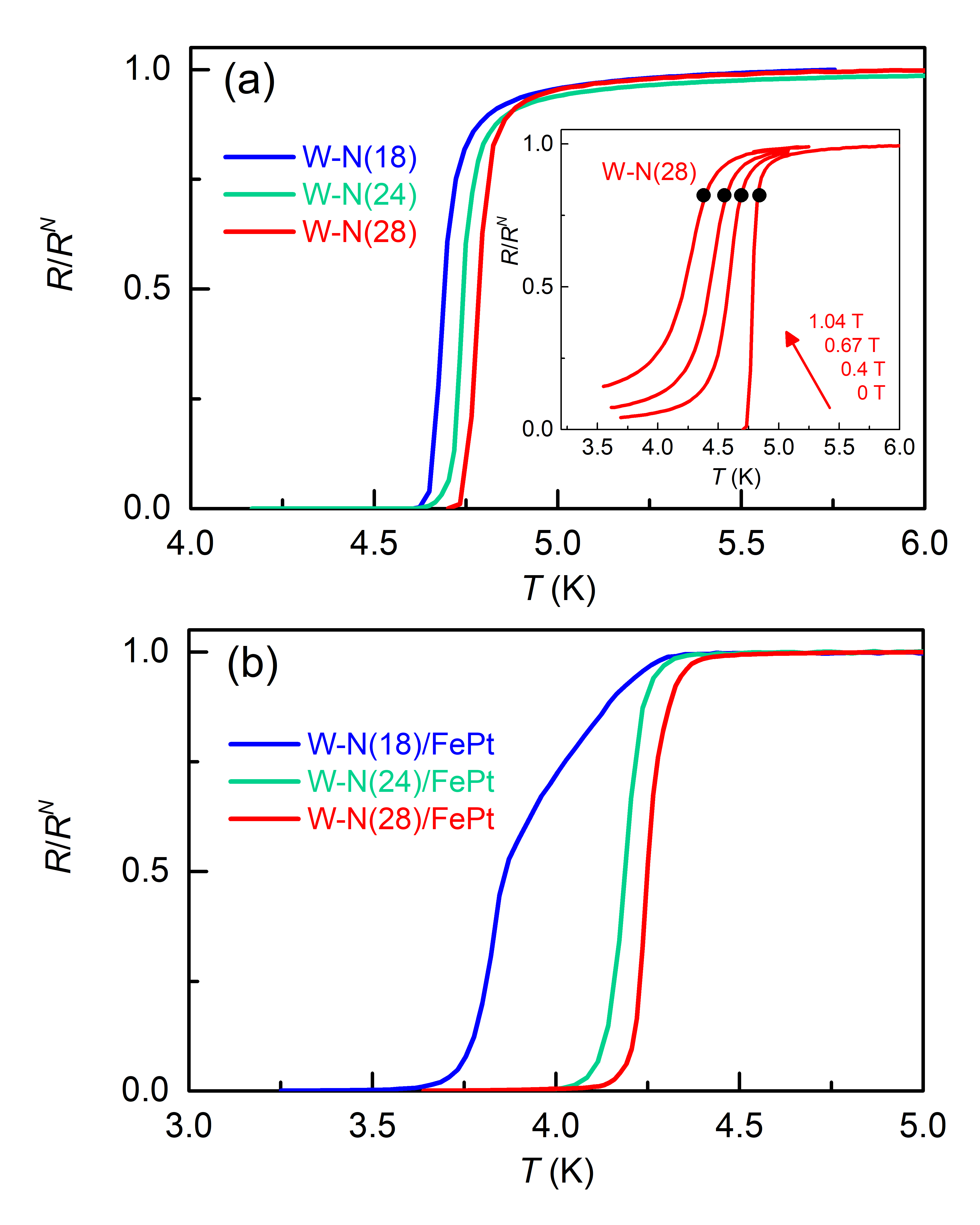}
\caption{Superconducting transitions of W--N single layers and W--N/FePt bilayers. (a) Normalized resistance $R/R^N$ as a function of temperature for W--N(18), W--N(24), and W--N(28). The inset shows the field evolution of the transition for W--N(28) under magnetic fields applied perpendicular to the film plane. The black dots indicate the $R/R^N=0.8$ criterion used to determine $T_c$. (b) Normalized resistance as a function of temperature for the corresponding W--N/FePt bilayers, showing an increasing suppression of $T_c$ with decreasing W--N thickness.}
\label{fig:3}
\end{figure}

\begin{table*}[t]
\centering
\caption{Superconducting and normal-state parameters of the W--N single-layer films.
$d$ is the nominal film thickness, $T_c$ the superconducting transition temperature,
$\rho^N$ the normal-state resistivity, $\mu_0H_{c2}(0)$ the extrapolated upper critical field,
$\xi(0)$ the zero-temperature coherence length,
$\left.d\mu_0H_{c2}/dT\right|_{T_c}$ the slope of the upper critical field near $T_c$,
$D$ the diffusion coefficient, $I_{\mathrm{dep}}$ the estimated depairing current,
and $I_c^{\mathrm{sf}}$ the experimentally measured self-field critical current
at 3 K, for a bridge width of $w \approx 4.5~\mu$m.}
\label{tab:WN_properties}

\setlength{\tabcolsep}{3.8pt}
\renewcommand{\arraystretch}{1.18}

\begin{tabular}{lcccccccccc}
\toprule
Sample &
$d$ &
$T_c$ &
$\rho^N$ &
$\mu_0H_{c2}(0)$ &
$\xi(0)$ &
$-\left.d\mu_0H_{c2}/dT\right|_{T_c}$ &
$D$ &
$I_{\mathrm{dep}}(0)$ &
$I_{\mathrm{dep}}(3\,\mathrm{K})$ &
$I_c^{\mathrm{sf}}(3\,\mathrm{K})$ \\
&
(nm) &
(K) &
($\mu\Omega\cdot\mathrm{cm}$) &
(T) &
(nm) &
(T\,K$^{-1}$) &
(cm$^2$\,s$^{-1}$) &
(mA) &
(mA) &
(mA) \\
\midrule
W--N(18) & 18 & 4.7  & 130(10) & 7.6(1) & 6.6 & 2.35 & 0.5 & 5.0 & 2.3 & -- \\
W--N(24) & 24 & 4.75 & 125(10) & 7.5(2) & 6.7 & 2.3 & 0.5 & 6.9 & 3.2 & 1.12 \\
W--N(28) & 28 & 4.8  & 115(10) & 7.3(2) & 6.7 & 2.2 & 0.5 & 8.8 & 4.2 & 1.55 \\
\bottomrule
\end{tabular}
\end{table*}

\subsection{Absence of edge-induced rectification in single-layer W–N microbridges}

Before examining the superconducting diode effect in W–N/FePt bilayers, we investigated the I--V characteristics of the single-layer W–N microbridges to assess possible contributions from geometric edge asymmetries. Previous studies have demonstrated that asymmetric edge indentations can generate a vortex-ratchet effect under an applied perpendicular magnetic field. In particular, Porrati et al.~\cite{Porrati2025} showed that edge indentations locally reduce the vortex-entry barrier, leading to different critical currents for opposite current polarities.

Figure~4 shows representative I--V characteristics of the single-layer W--N(28) microbridge measured at 3~K in perpendicular magnetic fields
up to 28~mT. At zero applied field, the bridge switches at $I \simeq \pm 1.55$~mA, giving a self-field critical current of about 37$\%$ of its estimated depairing current (4.2~mA). The single-layer W--N(24) bridge reaches a similar fraction at the same temperature, with a measured self-field critical current of 1.12~mA, or 35$\%$ of its estimated depairing current (3.2~mA, Table~\ref{tab:WN_properties}). For W--N(28), increasing the perpendicular field reduces the magnitude
of the switching currents and shifts the onset of dissipation to lower currents, consistent with vortex motion (Fig.~4, inset). The I--V
characteristics remain symmetric under current reversal within experimental uncertainty throughout the measured field range. The W--N(18) and W--N(24) bridges show the same symmetry (curves not shown). Thus, any vortex-ratchet effect associated with nanoscale edge irregularities such as those visible in Fig.~1(a) is below the resolution of these measurements.

\begin{figure}[t]
\centering
\includegraphics[width=\linewidth]{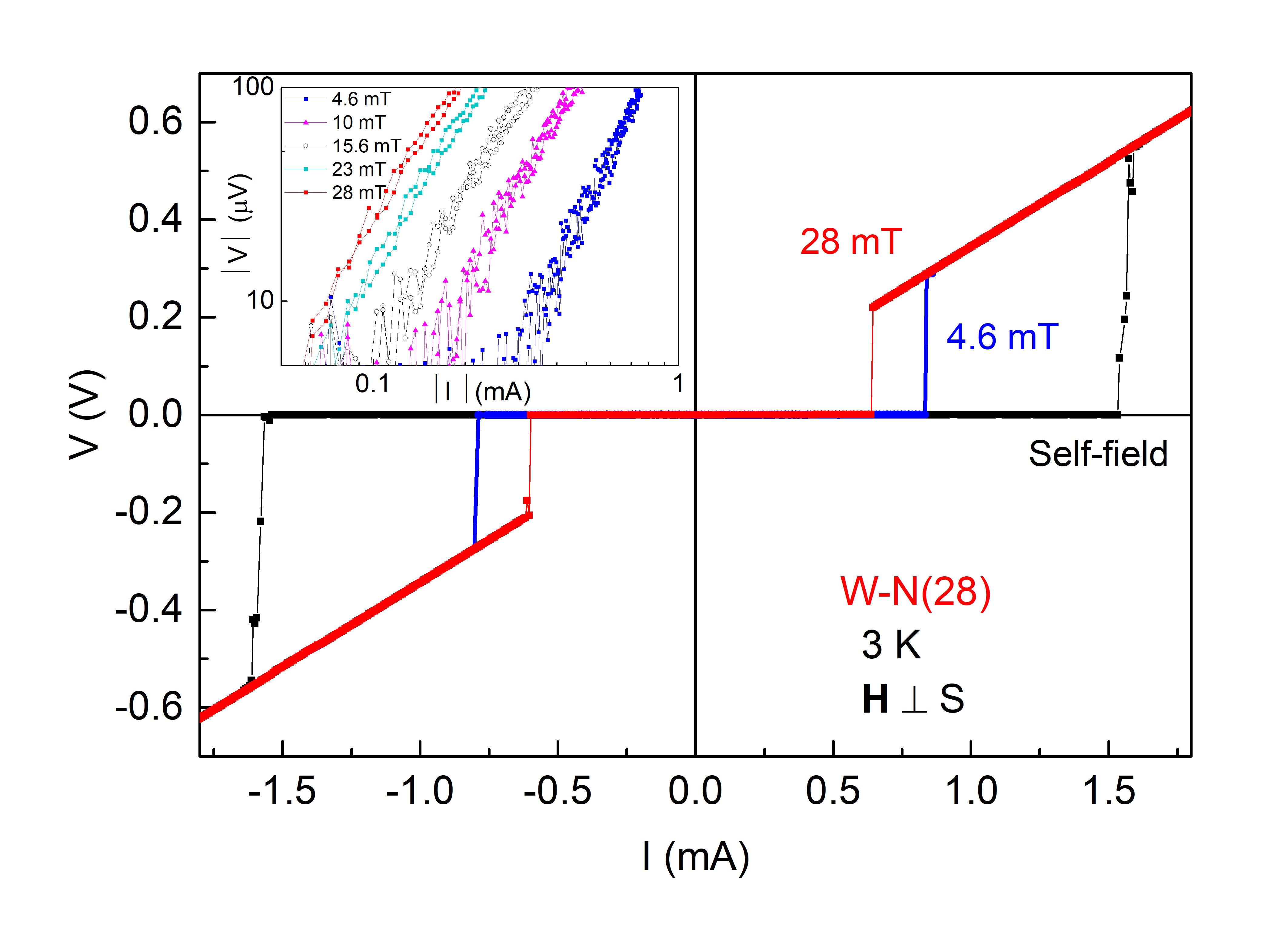}
\caption{Current--voltage (I--V) characteristics of the single-layer W--N(28) microbridge measured at 3 K under perpendicular magnetic fields. The main panel shows the nearly symmetric switching currents for positive and negative current polarities at zero field and at selected applied fields of 4.6 and 28 mT. The inset shows the evolution of the low-voltage dissipative regime with increasing magnetic field, plotted on logarithmic scales. The I--V characteristics remain symmetric within experimental uncertainty over the investigated field range, indicating no detectable rectification associated with geometric edge asymmetries.}
\label{fig:4}
\end{figure}

\subsection{Field-free diode effect and magnetic reconfiguration}

Previous results on Mo$_2$N/FePt and NbN/FePt heterostructures by Cruz et al.~\cite{Cruz2026} and Zapata et al.~\cite{Zapata2026NbNFePt} established a physical framework for understanding the field-free superconducting diode response in terms of the remanent magnetic configuration of FePt. Micromagnetic simulations revealed substantial perpendicular magnetic stray fields, with characteristic amplitudes of approximately 160--200 mT in the regular stripe regions and local enhancements reaching 280--365 mT near the microbridge edges. Although these values correspond to previously investigated FePt configurations, they provide a representative scale for the magnetic fields generated by this domain structure. In a thin superconducting film subject to a nanoscale alternating field, the local field amplitude alone is insufficient to determine whether spontaneous vortex--antivortex populations develop in the bridge interior. The simulations also identified a dipolar-like contribution associated with the extended in-plane remanent magnetization, producing opposite-signed perpendicular field corrections near the two bridge edges. Superimposed on the periodic stripe modulation, these contributions could favor preferential vortex and antivortex populations near opposite edges, providing a possible origin for the observed nonreciprocal transport.

\begin{figure}[t]
\centering
\includegraphics[width=\linewidth]{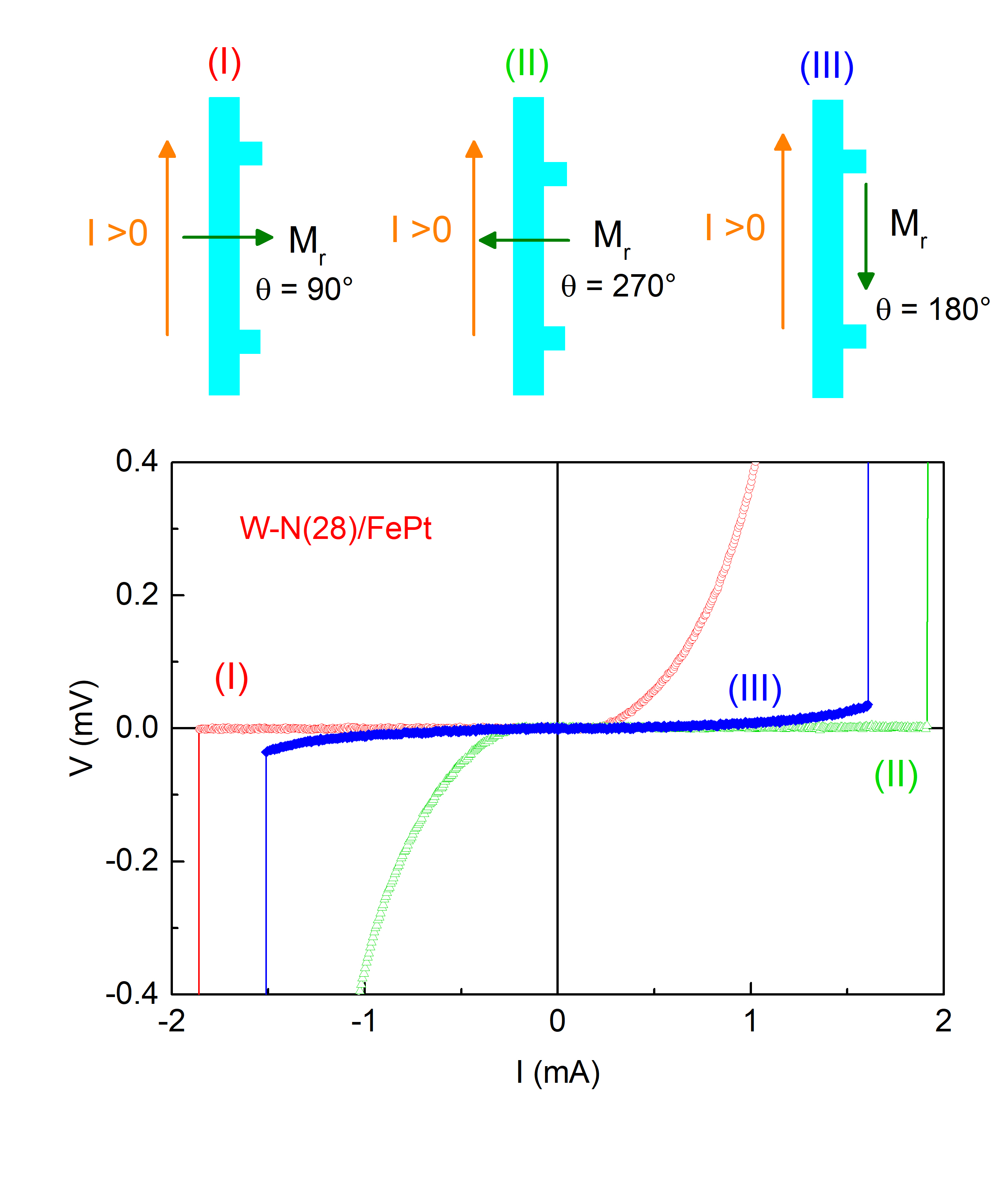}
\caption{Field-free I--V characteristics of W--N(28)/FePt at 3~K for three remanent FePt magnetization directions. The upper panels show the positive current direction and the remanent magnetization after an in-plane field of 0.5~T was applied and removed. The lower panel shows the corresponding I--V curves. Configurations~I (red, $\theta=90^\circ$) and~II (green, $\theta=270^\circ$) have opposite transverse magnetizations and opposite diode polarities. In
configuration~III (blue, $\theta=180^\circ$), the magnetization is antiparallel to the positive current direction and the response is nearly symmetric.}
\label{fig:5}
\end{figure}

Figure~\ref{fig:5} shows the zero-field I--V characteristics of W--N(28)/FePt at 3~K for the three remanent magnetic configurations illustrated in the upper panels. In configuration~I ($\theta=90^\circ$), the positive-current branch develops voltage gradually, with $I_c^+ \simeq +0.25$~mA at the $2~\mu\mathrm{V}$ criterion. The negative-current branch remains nearly dissipationless until it switches abruptly at $I_c^- \simeq -1.8$~mA. Reversing the transverse remanent
magnetization (configuration~II, $\theta=270^\circ$) interchanges the two branches and reverses the diode polarity. When the remanent magnetization is antiparallel to the positive current direction (configuration~III, $\theta=180^\circ$), the I--V characteristic becomes nearly symmetric. Together with the absence of detectable rectification in the single-layer W--N bridges (Fig.~4), these observations support a magnetic origin for the nonreciprocal response.

Within the edge-entry picture proposed for FePt-based bilayers~\cite{Cruz2026,Zapata2026NbNFePt}, the remanent stray field favors vortices at one lateral edge and antivortices at the other.
In configuration~I, positive current would drive both populations toward the bridge interior along the stripes, producing vortex motion transverse to the current.
For negative current, the Lorentz force would point toward the adjacent edge for each population, suppressing inward flux injection.
A longitudinal path near the center of the bridge could therefore remain depleted of mobile vortices up to switching.
In the ideal limit, this region would provide a vortex-free path for supercurrent. Transport is also compatible with vortices pinned within this region, for example at stripe defects.

\begin{figure*}[t!]
\centering
\includegraphics[width=\textwidth]{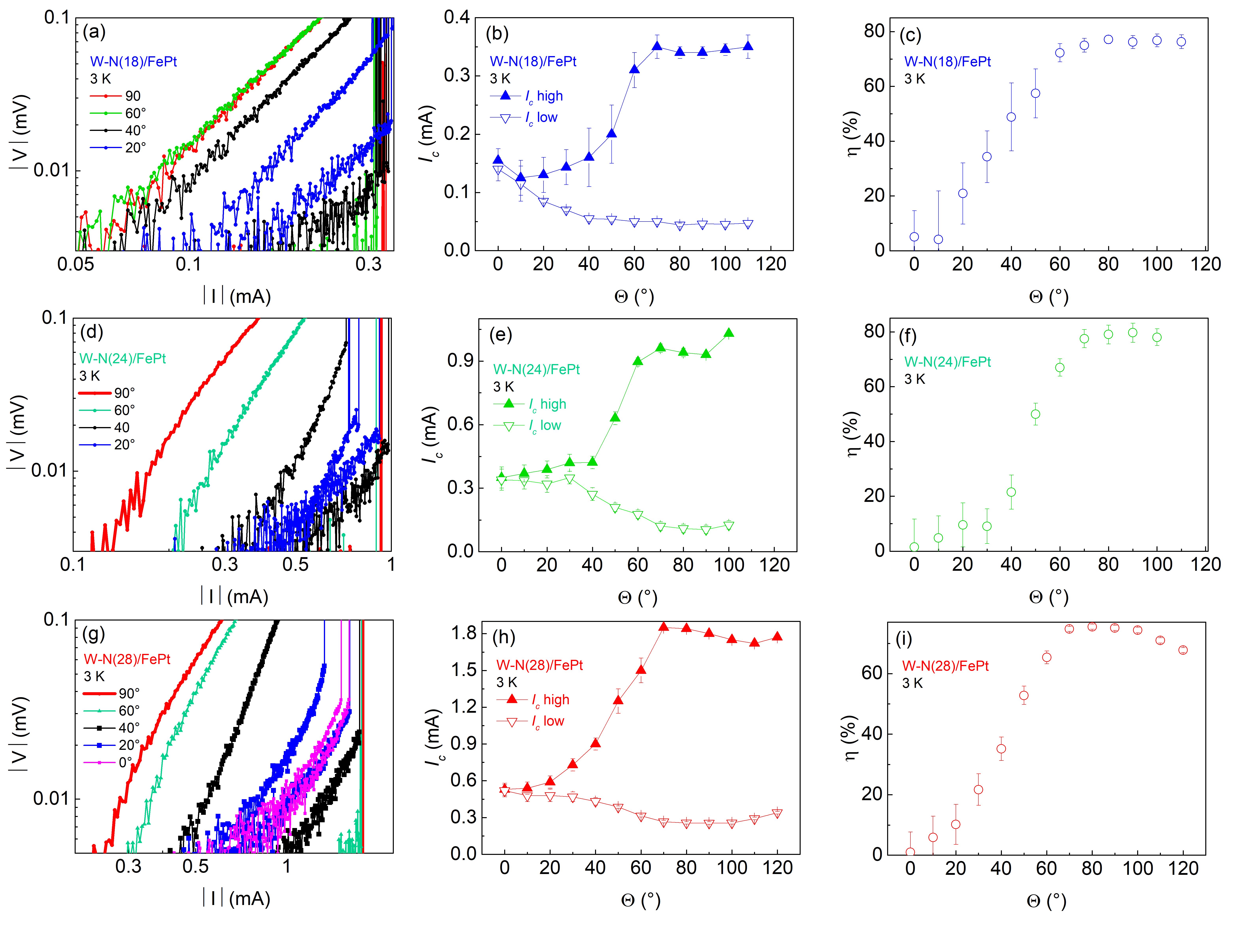}
\caption{Angular dependence of the field-free superconducting diode response in W--N/FePt bilayers at 3 K. (a--c) W--N(18)/FePt, (d--f) W--N(24)/FePt,
and (g--i) W--N(28)/FePt. (a,d,g) Absolute voltage $|V|$ as a function of absolute current $|I|$, plotted on logarithmic scales for selected magnetic preparation angles $\theta$. Vertical lines indicate the abrupt switching currents. Critical currents were determined using a voltage criterion of $|V|=2\,\mu\mathrm{V}$. For curves exhibiting a gradual onset of dissipation, $I_c$ was obtained by extrapolating the approximately linear portion of the log--log I--V characteristics to this voltage criterion. For abrupt transitions crossing this threshold, $I_c$ was identified with the switching current. (b,e,h) Angular dependence of the critical currents of the high- and low-current branches, $I_c^{\mathrm{high}}$ (filled symbols) and $I_c^{\mathrm{low}}$ (open symbols), respectively. (c,f,i) Corresponding diode efficiencies $\eta$, calculated using Eq.~\eqref{eq:eta}. The magnetic states were prepared by applying an in-plane saturation field of 0.5 T at an angle $\theta$ relative to the transport current and subsequently removing it. All I--V measurements were performed at zero applied magnetic field. The three bilayers exhibit similar angular dependencies, with maximum rectification near $\theta=90^\circ$ and strongly suppressed nonreciprocity toward $\theta=0^\circ$, despite their different operating-current scales. Error bars in (b,e,h) represent the experimental uncertainties in the critical currents, while those in (c,f,i) were obtained by uncertainty propagation.}
\label{fig:6}
\end{figure*}

Figure~\ref{fig:6} summarizes the angular dependence of the superconducting diode response at 3 K for the three W--N/FePt bilayers. The low-voltage I--V characteristics for W--N(18)/FePt, W--N(24)/FePt, and W--N(28)/FePt are shown in Figs.~\ref{fig:6}(a,d,g), respectively, for selected magnetic preparation angles. Despite their different operating-current scales, all three devices exhibit a similar evolution with the orientation of the remanent magnetic configuration. For preparation fields perpendicular to the current ($\theta=90^\circ$), a pronounced asymmetry develops between the two current polarities. One branch remains nearly dissipationless up to an abrupt switching transition, whereas the opposite branch develops a gradual voltage increase at substantially lower currents. As the preparation angle approaches the parallel configuration ($\theta=0^\circ$), the initially nearly dissipationless branch progressively develops a finite-voltage response before switching, reducing its critical current defined by the $2\,\mu$V criterion. Simultaneously, the onset of dissipation in the opposite branch shifts toward higher currents, progressively reducing the asymmetry between the two dissipation thresholds.

The corresponding critical-current magnitudes, $I_c^{\mathrm{high}}$ and $I_c^{\mathrm{low}}$, are shown in Figs.~\ref{fig:6}(b,e,h). Their opposite angular trends are consistent with changes in the conditions for vortex entry and subsequent motion within the magnetic stripe structure. When the stripes are oriented approximately perpendicular to the current, the Lorentz force acts predominantly along the stripe-defined magnetic channels, potentially favoring vortex motion for the polarity that drives flux inward~\cite{Cruz2026,Zapata2026NbNFePt}. Rotating the stripes toward the current direction increases the component of the Lorentz force across the magnetic modulation, where barriers to transverse vortex motion may increase $I_c^{\mathrm{low}}$. At the same time, a reduction in the edge-field asymmetry may facilitate the onset of dissipation in the opposite polarity, reducing $I_c^{\mathrm{high}}$ without necessarily producing a comparable reduction in its abrupt switching current. This distinction is particularly evident in the three samples. For example, in W--N(28)/FePt at $\theta=0^\circ$, both branches reach the 2 $\mu$V criterion at approximately 0.5 mA, whereas abrupt switching occurs at substantially higher currents, around 1.8 mA [Fig.~\ref{fig:6}(g,h)].

The diode efficiency is defined as
\begin{equation}
  \eta = \frac{\left|\,|I_c^+|-|I_c^-|\,\right|}
  {|I_c^+|+|I_c^-|},
  \label{eq:eta}
\end{equation}
where $I_c^+$ and $I_c^-$ are the critical currents for positive and negative current polarities, respectively. The resulting diode efficiencies are presented in Figs.~6(c), 6(f), and 6(i). All three bilayers show a similar angular dependence, with a gradual increase in efficiency as the magnetic preparation angle changes from parallel toward perpendicular to the current. Maximum efficiencies of approximately 75--80\% are obtained near the perpendicular configuration, corresponding to critical-current ratios
$I_c^{\mathrm{high}}/I_c^{\mathrm{low}}$ of approximately 7--9. Thus, the two critical currents differ by nearly an order of magnitude
in all three devices. In contrast, the rectification is strongly suppressed as the preparation angle approaches $0^\circ$. 

At 3~K and $\theta=90^\circ$, $I_c^{\mathrm{high}}$ is approximately 0.35, 0.95, and 1.8~mA for W--N layer thicknesses of 18, 24, and 28~nm, respectively. The corresponding values of $I_c^{\mathrm{low}}$ are about 0.05, 0.10, and 0.25~mA.
The corresponding current windows between the two onset thresholds are approximately 0.30, 0.85, and 1.55~mA.
For 24 and 28~nm, the high-current thresholds are comparable to the self-field critical currents of the corresponding single-layer bridges (1.12 and 1.55~mA in Table~\ref{tab:WN_properties}). The low-current thresholds are only about one tenth and one sixth of those reference currents, respectively.
The increase in operating current accompanies both the increase in cross section and the smaller suppression of $T_c$ in the thicker bilayers [Fig.~\ref{fig:3}(b)].

\subsection{Operating-current windows and resistive switching}

Beyond the diode efficiency, an important figure of merit of the present devices is the accessible current window over which opposite current polarities correspond to different transport states. Figure~7(a) illustrates this behavior for W--N(28)/FePt at 3 and 3.75 K, under the remanent magnetic configuration corresponding to maximum rectification. In the low-current branch, dissipation develops gradually above $I_c^{\mathrm{low}}$, consistent with vortex motion. At higher currents, this dissipative branch undergoes an abrupt switching transition at $I_c^{*}$. In contrast, the high-current branch remains nearly dissipationless up to a much larger current $I_c^{\mathrm{high}}$, where an abrupt transition to the resistive state occurs. These characteristics allow the device to operate in different rectification regimes depending on the applied current.

The overall operating-current window is defined as
\begin{equation}
\Delta I_{\mathrm{S-D}} =
I_c^{\mathrm{high}}-I_c^{\mathrm{low}},
\end{equation}
corresponding to the range of current magnitudes over which only one polarity exceeds the $2\,\mu\mathrm{V}$ criterion. Within this interval, two distinct regimes can be identified. Between $I_c^{\mathrm{low}}$ and $I_c^{*}$, dissipation is associated with the gradual voltage development characteristic of vortex motion. At $I_c^{*}$, the dissipative branch undergoes an abrupt transition to a high-resistance state. This defines a second, more restrictive operating window,
\begin{equation}
\Delta I_{\mathrm{S-N}} =
I_c^{\mathrm{high}}-I_c^{*},
\end{equation}
in which one current polarity remains superconducting while the opposite polarity has already switched to the high-resistance state. At 3~K, $I_c^{*}\simeq1.05$~mA for W--N(28)/FePt, giving $\Delta I_{\mathrm{S-D}}\simeq1.55$~mA and $\Delta I_{\mathrm{S-N}}\simeq0.75$~mA. One possible origin of the abrupt switching at $I_c^{*}$ is a Larkin--Ovchinnikov-type flux-flow instability~\cite{Larkin1975}. The remanent FePt stray fields can supply flux for vortex motion at zero applied field, and the transport self-field may also contribute. The observed voltage jump is also compatible with a thermal contribution.

\begin{figure}[t!]
\centering
\includegraphics[width=\linewidth]{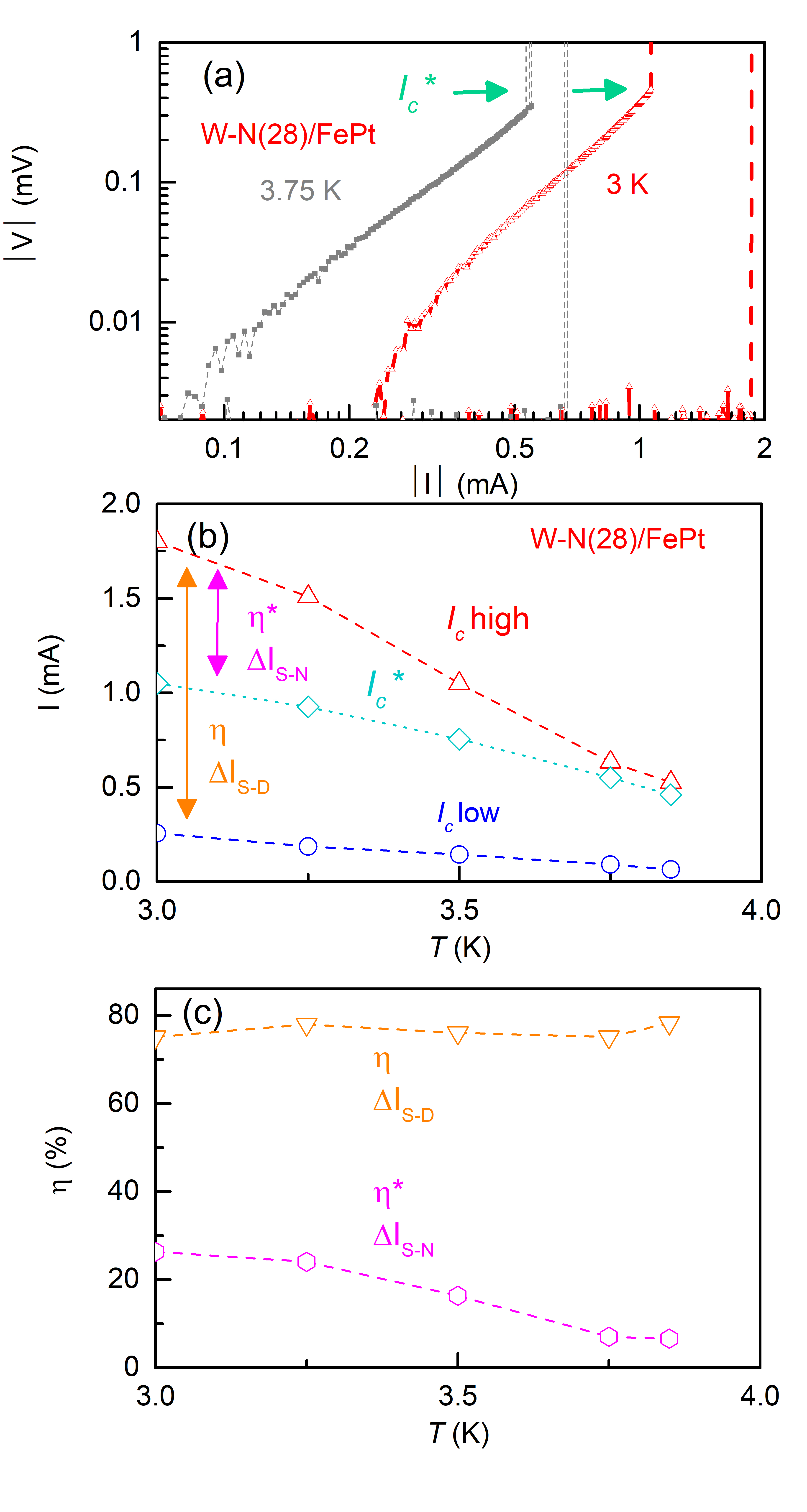}
\caption{Temperature dependence of the operating-current windows and rectification efficiencies in W--N(28)/FePt under the remanent magnetic configuration corresponding to maximum diode response. (a) Low-voltage I--V characteristics at 3 and 3.75~K, plotted as magnitudes for both polarities, showing the gradual development of dissipation in the low-current branch and the nearly dissipationless high-current branch. Vertical dashed lines indicate the abrupt switching currents. (b) Temperature dependence of the high- and low-branch critical currents, $I_c^{\mathrm{high}}$ and $I_c^{\mathrm{low}}$, together with the switching current $I_c^{*}$ of the dissipative branch. The orange and magenta arrows indicate the current differences entering the numerators of $\eta$ and $\eta^{*}$, respectively. (c) Corresponding temperature dependence of the conventional diode efficiency $\eta$, calculated using Eq.~\eqref{eq:eta}, and the efficiency $\eta^{*}$ associated with the high-resistance switching window, defined in Eq.~\eqref{eq:eta_star}. All measurements were performed at zero applied magnetic field after preparing the remanent state of FePt.}
\label{fig:7}
\end{figure}

Figure~\ref{fig:7}(b) shows that all three characteristic currents decrease with increasing temperature in W--N(28)/FePt. Both operating windows narrow in absolute current. To compare how the widths change relative to the current scale, we use $\eta$ from Eq.~\eqref{eq:eta} and define a second efficiency from the switching current $I_c^{*}$,
\begin{equation}
  \eta^{*} = \frac{\left|\,|I_c^{\mathrm{high}}|-|I_c^{*}|\,\right|}
  {|I_c^{\mathrm{high}}|+|I_c^{*}|},
  \label{eq:eta_star}
\end{equation}
From 3 to 3.85~K, the ratio $I_c^{\mathrm{high}}/I_c^{\mathrm{low}}$ varies little, and $\eta$ stays near 75--80\% [Fig.~\ref{fig:7}(c)]. Over the same range, $I_c^{*}/I_c^{\mathrm{high}}$ increases and $\eta^{*}$ falls from about 25\% to about 7\%. The switching-based window therefore narrows both in absolute current and as a fraction of the operating-current scale.

These results demonstrate that the operating characteristics of W--N/FePt diodes can be adjusted through several complementary parameters. Increasing the W--N thickness provides access to higher operating currents, although the resulting scale also depends on proximity-induced suppression of superconductivity and the corresponding transition temperature. In W--N(28)/FePt, temperature also tunes the operating currents while the conventional diode efficiency remains nearly constant. Furthermore, selecting the operating-current interval allows rectification between superconducting and dissipative states or between superconducting and the normal-resistance state, with different effective efficiencies. In the latter regime, the resistance and voltage response could, in principle, be further adjusted by modifying the bridge length while maintaining the same cross section. This geometrical degree of freedom provides a possible route to tailoring the output voltage of the devices for different cryogenic circuit applications.

\section{Summary and conclusions}

We have demonstrated high-efficiency, field-free superconducting diode behavior in amorphous W--N/FePt bilayers fabricated entirely by sputtering at room temperature. Across the three W--N thicknesses investigated, diode efficiencies of approximately 75--80$\%$ are obtained at 3 K without intentionally asymmetric device geometries. The response is controlled by the remanent magnetic configuration of FePt: it is maximized when the magnetic preparation field is perpendicular to the transport current, strongly suppressed for parallel alignment, and reversed upon reversal of the remanent magnetization.

At 3~K and $\theta=90^\circ$, increasing the W--N thickness from 18 to 28~nm raises $I_c^{\mathrm{high}}$ from approximately 0.35 to 1.8~mA and widens the interval between the two dissipation thresholds by a factor of about five. This trend reflects the combined influence of the increasing superconducting cross section and the thickness-dependent suppression of $T_c$ in the bilayers. Increasing temperature provides an additional means of reducing the operating-current scale while preserving a nearly constant conventional diode efficiency. This behavior is observed across the three W--N/FePt bilayers and is illustrated for W--N(28)/FePt in Fig.~\ref{fig:7}.

The angular dependence and reversal of the diode polarity support a magnetic origin of the nonreciprocal response, consistent with the edge-controlled vortex-entry mechanism proposed for FePt-based bilayers \cite{Cruz2026,Zapata2026NbNFePt}. In this picture, the remanent stray fields may favor vortex and antivortex populations near opposite bridge edges. Depending on the current polarity, the Lorentz force can either promote their injection and guided motion along the magnetic stripes, producing dissipation, or drive them outward, suppressing the flow of mobile vortices through the bridge. Importantly, the nearly dissipationless branch does not necessarily imply a vortex-free interior. Vortices may remain strongly pinned at local defects or magnetic inhomogeneities without participating in guided flux flow. Although the microscopic vortex distribution cannot be established from transport measurements alone, this scenario provides a consistent interpretation of the observed polarity-dependent dissipation.

The coexistence of gradual dissipation and abrupt resistive switching defines distinct operating windows. The conventional diode efficiency, $\eta \approx 75$--$80\%$, characterizes the asymmetry of the dissipation-onset currents. By comparison, the switching-based efficiency, $\eta^* \approx 25\%$ at 3 K for W--N(28)/FePt, characterizes the narrower current window in which one polarity remains nearly dissipationless while the opposite polarity has switched to the normal-resistance state. This distinction connects the reported efficiencies to the transport states available for rectification. Superconducting thickness, temperature, and magnetic preparation thus provide complementary means of setting the operating-current range and polarity of field-free SC/FePt rectifiers.

\section*{CRediT authorship contribution statement}
N. Haberkorn conceived the project. S. Russo, A. Sanson, and N. Haberkorn fabricated the samples. S. Russo and A. Sanson performed the electrical transport measurements. Y. Lee and J. Kim performed MFM measurements. All authors discussed the results and contributed to writing and revising the manuscript. All authors read and approved the final version of the manuscript.

\section*{Acknowledgements}
The authors acknowledge S. Anguiano and C. Bertoli for technical assistance. This work was partially supported by the CONICET (PIP 11220210100263CO), BrainLink program funded by the Ministry of Science and ICT through the National Research Foundation of Korea (No. RS-2024-00410027 and No. RS-2026-25472018) and Brain Pool program funded by the Ministry of Science and ICT through the National Research Foundation of Korea (RS-2023-00222408). 

\section*{Declaration of competing interest}
The authors declare no competing interests.

\section*{Data availability}
The data supporting the findings of this study are available from the corresponding author upon reasonable request.

\bibliographystyle{elsarticle-num}
\bibliography{references}

@article{JiangHu2022,
  author  = {Jiang, Kun and Hu, Jiangping},
  title   = {Superconducting diode effects},
  journal = {Nat. Phys.},
  volume  = {18},
  pages   = {1145--1146},
  year    = {2022},
  doi     = {10.1038/s41567-022-01701-0}
}

@article{Nadeem2023,
  author  = {Nadeem, M. and Fuhrer, M. S. and Wang, X.},
  title   = {The Superconducting Diode Effect},
  journal = {Nat. Rev. Phys.},
  volume  = {5},
  pages   = {558--577},
  year    = {2023},
  doi     = {10.1038/s42254-023-00632-w}
}

@article{Golod2022,
  author  = {Golod, Taras and Krasnov, Vladimir M.},
  title   = {Demonstration of a superconducting diode-with-memory, operational at zero magnetic field with switchable nonreciprocity},
  journal = {Nat. Commun.},
  volume  = {13},
  pages   = {3658},
  year    = {2022},
  doi     = {10.1038/s41467-022-31256-w}
}

@article{Le2024,
  author  = {Le, Tian and Pan, Zhiming and Xu, Zhuokai and Liu, Jinjin and Wang, Jialu and Lou, Zhefeng and Yang, Xiaohui and Wang, Zhiwei and Yao, Yugui and Wu, Congjun and Lin, Xiao},
  title   = {Superconducting diode effect and interference patterns in kagome {CsV$_3$Sb$_5$}},
  journal = {Nature},
  volume  = {630},
  pages   = {64--69},
  year    = {2024},
  doi     = {10.1038/s41586-024-07431-y}
}

@article{Lin2022,
  author  = {Lin, Jiang-Xiazi and Siriviboon, Phum and Scammell, Harley D. and Liu, Song and Rhodes, Daniel and Watanabe, K. and Taniguchi, T. and Hone, James and Scheurer, Mathias S. and Li, J. I. A.},
  title   = {Zero-field superconducting diode effect in small-twist-angle trilayer graphene},
  journal = {Nat. Phys.},
  volume  = {18},
  pages   = {1221--1227},
  year    = {2022},
  doi     = {10.1038/s41567-022-01700-1}
}

@article{Hou2023,
  author  = {Hou, Y. and Nichele, F. and Chi, H. and Lodesani, A. and Wu, Y. and Ritter, M. F. and Haxell, D. Z. and Davydova, M. and Ili\'{c}, S. and Glezakou-Elbert, O. and Varambally, A. and Bergeret, F. S. and Kamra, A. and Fu, L. and Lee, P. A. and Moodera, J. S.},
  title   = {Ubiquitous Superconducting Diode Effect in Superconductor Thin Films},
  journal = {Phys. Rev. Lett.},
  volume  = {131},
  pages   = {027001},
  year    = {2023},
  doi     = {10.1103/PhysRevLett.131.027001}
}

@article{Lyu2021,
  author  = {Lyu, Yang-Yang and Jiang, Ji and Wang, Yong-Lei and Xiao, Zhi-Li and Dong, Sining and Chen, Qing-Hu and Milo{\v{s}}evi{\'c}, Milorad V. and Wang, Huabing and Divan, Ralu and Pearson, John E. and Wu, Peiheng and Peeters, Francois M. and Kwok, Wai-Kwong},
  title   = {Superconducting diode effect via conformal-mapped nanoholes},
  journal = {Nat. Commun.},
  volume  = {12},
  pages   = {2703},
  year    = {2021},
  doi     = {10.1038/s41467-021-23077-0}
}

@article{Porrati2025,
  author  = {Porrati, F. and Pokusinskyi, A. O. and Barth, S. and Huth, M. and Dobrovolskiy, O. V.},
  title   = {Vortex Ratchet Effect in a {NbC} Strip With a Periodic Edge Indentation},
  journal = {Small Methods},
  volume  = {9},
  number  = {12},
  pages   = {e01430},
  year    = {2025},
  doi     = {10.1002/smtd.202501430}
}

@article{Carapella2010,
  author  = {Carapella, G. and Sabatino, P. and Costabile, G.},
  title   = {Asymmetry, bistability, and vortex dynamics in a finite-geometry ferromagnet-superconductor bilayer structure},
  journal = {Phys. Rev. B},
  volume  = {81},
  pages   = {054503},
  year    = {2010},
  doi     = {10.1103/PhysRevB.81.054503}
}

@article{Zapata2026,
  author  = {Zapata, J. C. and De Biasi, E. and Sirena, M. and Kim, J. and Haberkorn, N.},
  title   = {Field-Free Memory-Programmable Superconducting Diode},
  journal = {Adv. Funct. Mater.},
  volume  = {36},
  pages   = {e75836},
  year    = {2026},
  doi     = {10.1002/adfm.75836}
}

@article{Vodolazov2005,
  author  = {Vodolazov, D. Y. and Peeters, F. M.},
  title   = {Superconducting rectifier based on the asymmetric surface barrier effect},
  journal = {Phys. Rev. B},
  volume  = {72},
  pages   = {172508},
  year    = {2005},
  doi     = {10.1103/PhysRevB.72.172508}
}

@article{Aladyshkin2010,
  author  = {Aladyshkin, A. Y. and Vodolazov, D. Y. and Fritzsche, J. and Kramer, R. B. G. and Moshchalkov, V. V.},
  title   = {Reverse-Domain Superconductivity in Superconductor--Ferromagnet Hybrids: Effect of a Vortex-Free Channel on the Symmetry of {I--V} Characteristics},
  journal = {Appl. Phys. Lett.},
  volume  = {97},
  pages   = {052501},
  year    = {2010},
  doi     = {10.1063/1.3474622}
}

@article{Cerbu2013,
  author  = {Cerbu, D. and Gladilin, V. N. and Cuppens, J. and Fritzsche, J. and Tempere, J. and Devreese, J. T. and Moshchalkov, V. V. and Silhanek, A. V. and Van de Vondel, J.},
  title   = {Vortex ratchet induced by controlled edge roughness},
  journal = {New J. Phys.},
  volume  = {15},
  pages   = {063022},
  year    = {2013},
  doi     = {10.1088/1367-2630/15/6/063022}
}

@misc{Cruz2026,
  author        = {Cruz, Juan S. and De Biasi, Emilio and Zapata, Juan C. and Lee, Yeonkyu and Kim, Jeehoon and Haberkorn, Nestor},
  title         = {High-Efficiency Field-Free Superconducting Diodes in {Mo$_2$N}/{FePt} Bilayers},
  year          = {2026},
  howpublished  = {arXiv:2610.16972},
  eprint        = {2610.16972},
  archivePrefix = {arXiv},
  primaryClass  = {cond-mat.supr-con},
  doi           = {10.48550/arXiv.2610.16972}
}

@misc{Zapata2026NbNFePt,
  author        = {Zapata, Juan C. and De Biasi, Emilio and Sirena, Martin and Kim, Jeehoon and Haberkorn, Nestor},
  title         = {Field-free superconducting diode and tunable vortex dynamics in {NbN}/{FePt} stripe-domain heterostructures},
  year          = {2026},
  howpublished  = {arXiv:2610.30556},
  eprint        = {2610.30556},
  archivePrefix = {arXiv},
  primaryClass  = {cond-mat.supr-con},
  doi           = {10.48550/arXiv.2610.30556}
}

@article{Voltan2016,
  author  = {Voltan, S. and Cirillo, C. and Snijders, H. J. and Lahabi, K. and Garc{\'i}a-Santiago, A. and Hern{\'a}ndez, J. M. and Attanasio, C. and Aarts, J.},
  title   = {Emergence of the stripe-domain phase in patterned permalloy films},
  journal = {Phys. Rev. B},
  volume  = {94},
  pages   = {094406},
  year    = {2016},
  doi     = {10.1103/PhysRevB.94.094406}
}

@article{Caputo2017,
  author  = {Caputo, M. and Cirillo, C. and Voltan, S. and Cucolo, A. M. and Aarts, J. and Attanasio, C.},
  title   = {Influence of the magnetic configuration on the vortex-lattice instability in {Nb}/permalloy bilayers},
  journal = {Phys. Rev. B},
  volume  = {96},
  pages   = {174519},
  year    = {2017},
  doi     = {10.1103/PhysRevB.96.174519}
}

@article{Alvarez2015,
  author  = {{\'{A}lvarez}, N. R. and V{\'a}zquez Montalbetti, M. E. and G{\'o}mez, J. E. and Moya Riffo, A. E. and Vicente {\'{A}lvarez}, M. A. and Goovaerts, E. and Butera, A.},
  title   = {Tunable stress induced magnetic domain configuration in {FePt} thin films},
  journal = {J. Phys. D: Appl. Phys.},
  volume  = {48},
  pages   = {405003},
  year    = {2015},
  doi     = {10.1088/0022-3727/48/40/405003}
}

@article{Blatter2024,
  author  = {Blatter, G. and Sirena, M. and Lee, Yeonkyu and Yun, Jinyoung and Kim, Jeehoon and De Biasi, E. and Haberkorn, N.},
  title   = {Influence of thickness and domain structure on the vortex instability of superconducting/ferromagnetic bilayers},
  journal = {Supercond. Sci. Technol.},
  volume  = {37},
  number  = {3},
  pages   = {035007},
  year    = {2024},
  doi     = {10.1088/1361-6668/ad221f}
}

@article{Hofer2023,
  author  = {Hofer, J. A. and Bengio, S. and Su\'{a}rez, S. and Haberkorn, N.},
  title   = {Elucidating the role of disorder introduced by nitrogen in the superconducting properties of tungsten thin films},
  journal = {Mater. Adv.},
  volume  = {4},
  pages   = {150--156},
  year    = {2023},
  doi     = {10.1039/D2MA00935H}
}

@article{Hofer2019,
  author  = {Hofer, J. A. and Haberkorn, N.},
  title   = {Superconductivity in nanocrystalline tungsten thin films growth by sputtering in a nitrogen-argon mixture},
  journal = {Thin Solid Films},
  volume  = {685},
  pages   = {117--122},
  year    = {2019},
  doi     = {10.1016/j.tsf.2019.06.014}
}

@article{Zapata2025MatLett,
  author  = {Zapata, J. C. and Kim, Jeehoon and Haberkorn, N.},
  title   = {Geometric scaling of vortex critical velocities in {NbN} microbridges},
  journal = {Mater. Lett.},
  volume  = {382},
  pages   = {137945},
  year    = {2025},
  doi     = {10.1016/j.matlet.2024.137945}
}

@article{Werthamer1966,
  author  = {Werthamer, N. R. and Helfand, E. and Hohenberg, P. C.},
  title   = {Temperature and Purity Dependence of the Superconducting Critical Field, $H_{c2}$. {III}. Electron Spin and Spin-Orbit Effects},
  journal = {Phys. Rev.},
  volume  = {147},
  pages   = {295--302},
  year    = {1966},
  doi     = {10.1103/PhysRev.147.295}
}

@book{Tinkham1996,
  author    = {Tinkham, Michael},
  title     = {Introduction to Superconductivity},
  edition   = {2},
  publisher = {McGraw-Hill},
  address   = {New York},
  year      = {1996}
}

@article{ClemKogan2012,
  author  = {Clem, John R. and Kogan, V. G.},
  title   = {Kinetic impedance and depairing in thin and narrow superconducting films},
  journal = {Phys. Rev. B},
  volume  = {86},
  pages   = {174521},
  year    = {2012},
  doi     = {10.1103/PhysRevB.86.174521}
}

@article{Buzdin2005,
  author  = {Buzdin, A. I.},
  title   = {Proximity effects in superconductor-ferromagnet heterostructures},
  journal = {Rev. Mod. Phys.},
  volume  = {77},
  number  = {3},
  pages   = {935--976},
  year    = {2005},
  doi     = {10.1103/RevModPhys.77.935}
}

@article{Larkin1975,
  author  = {Larkin, A. I. and Ovchinnikov, Yu. N.},
  title   = {Nonlinear conductivity of superconductors in the mixed state},
  journal = {Sov. Phys. JETP},
  volume  = {41},
  number  = {5},
  pages   = {960--965},
  year    = {1975}
}

\end{document}